\documentclass[
aps,%
12pt,%
final,%
notitlepage,%
oneside,%
onecolumn,%
nobibnotes,%
nofootinbib,% 
superscriptaddress,%
noshowpacs,%
centertags]%
{revtex4}
\usepackage{epstopdf}

\begin{document}
\selectlanguage{english}

Astrophysical Bulletin, 2015, Vol. 70, No. 3, pp. 243--248.

\title{Photometric and Photopolarimetric Observations of a New Polar USNO-A2.0\,0825-18396733}%  §¡¨¥­¨¥ ­  áâப¨ ®áãé¥á⢫ï¥âáï ª®¬ ­¤®© \\

\author{\firstname{V.~L.}~\surname{Afanasiev}}
\affiliation{Special Astrophysical Observatory of Russian Academic of Sciences}
\author{\firstname{N.~V.}~\surname{Borisov}}
\affiliation{Special Astrophysical Observatory of Russian Academic of Sciences}
\author{\firstname{M.~M.}~\surname{Gabdeev}}\email{gamak@sao.ru}
\affiliation{Special Astrophysical Observatory of Russian Academic of Sciences}
\received{December 9, 2014}  \revised{May 6, 2015}

%\date{\today}
%\today ¯¥ç â ¥â c¥£®¤­ïè­¥¥ ç¨á«®

\begin{abstract}
We present photometric and polarimetric observations of a new
magnetic cataclysmic variable, presumably a polar,
USNO-A2.0\,0825-18396733. The photometric observations were
carried out with the SAO~RAS Zeiss-1000 telescope, and the
polarimetry was performed with the 6-m BTA telescope. The refined
orbital period $P=0.^d08481(2)$; the brightness of the system
varied from $17.5$ to $20.0$ in the $R_c$ filter. Polarimetric
observations of the object  in the $V$ filter revealed  strong
variable circular polarization: $-2\%$ to over $-31\%$. This
indicates that the variable USNO-A2.0\,0825-18396733 is a polar.
\end{abstract}

\maketitle

\section{INTRODUCTION}

 AM\,Her-type stars, or polars, are a subclass of cataclysmic variables.
 The system consists of a white dwarf
(the main component) and a red dwarf star that fills its Roche
lobe (the secondary component). Peculiar properties of polars are
associated with a strong magnetic field of the white dwarf. It
prevents the formation of an accretion disk and directs the
material along the magnetic field lines to the magnetic poles of
the white dwarf. The accretion column is one of the main sources
of radiation in the optical and X-ray ranges. The optical
radiation of polars is strongly polarized. Surveys  on the polars
are compiled by Voikhanskaya~\cite{voih1:Afanasiev_n} and
Cropper~\cite{crop1:Afanasiev_n}.

For the first time, strong polarization was found in AM\,Her, the
first representative of this type of systems.
Tapia~\cite{tapi1:Afanasiev_n} discovered linear and circular
polarization  in the $V$ and $I$ bands. Linear polarization
reached $6.8\%$ at the maximum, and the circular ranged from $4\%$
to $-9.5\%$. As a result of these observations, it was concluded
that the white dwarf has a strong magnetic
 field of about  $2\times10^8$~G. AM\,Her was later observed in its
low brightness
state~\mbox{\cite{shmi1:Afanasiev_n,lath1:Afanasiev_n,wick1:Afanasiev_n}},
and the white dwarf magnetic field strength as determined by the
locations of the Zeeman components of  hydrogen lines proved to be
$13$~MG, which is much lower than that proposed by Tapia. The
  magnetic field strength estimate based on  the degree of linear and circular
polarizations is inefficient, since they largely depend on the
geometry of a system. At the moment, the detection of polarization
of optical radiation is the main criterion for discovery of
polars.

The studied object USNO-A2.0\,0825-\linebreak 18396733 (hereafter
USNO\,0825) was found while   a region in the Aquila constellation
was observed by a team of amateur
astronomers~\cite{krya1:Afanasiev_n} using the  \mbox{30-cm}
Astrotel-Caucasus Observatory Ritchey--Chretien telescope
 and an Apogee Alta~U9000 CCD detector. They have obtained
98 images with 300-second exposures, without a filter.
 The system ephemeris was determined thereon:
\begin{equation*}
\begin{array}{r}
\label{shiequ1:Afanasiev_n}
\mbox{HJD}  =2455387.3976 (\pm 0.001) \\
        + 0.0840 (\pm 0.0004)\lefteqn{\,\mbox{E},}
\end{array}
\end{equation*}
where the zero phase corresponds to the time of minimum
brightness. A short period and a significant amplitude of the
brightness variability of the object are fairly typical of
cataclysmic variables. Strong  lines of H, He\,I, and the
He\,II\,4686\,\AA\ line are visible in the spectrum obtained by
N.~Borisov and V.~Shimansky on the 6-m telescope in August 2010.
The intensity ratio $I_{\rm He\,II\,4686}/I_{\rm H\beta} \sim
0.8$, which is typical of polars~\cite{voih2:Afanasiev_n}.

We have carried out photometric observations of USNO\,0825 to
clarify the period of variability, and the first polarimetric
observations to determine the nature of the object. Section~1
 describes the observations, Section~2 is devoted to the analysis of  the
observations, and  the conclusions of the study are given at the
end.

\section{OBSERVATIONS}
\subsection{Photometry}

Photometric observations were carried out on the SAO RAS 1-m
Zeiss-1000 telescope with the use of the standard photometer and a
CCD detector EEV\,\mbox{42-40} ($2048\!\times\!2048$~pixels sized
\mbox{$13.5\times13.5$~m}) with nitrogen cooling. The
observations were carried out in the  $R_c$ filter in autumn 2010,
exposure time was selected depending on weather conditions and
object brightness. A reference star and two comparison stars were
selected close to the object (Fig.~\ref{fig1:Afanasiev_n}). The
reference star was linked to standard stars 1925 ($V=12.39$,
$V-R=0.221$) and 2093 (\mbox{$V=12.54$}, \mbox{$V-R=0.370$}) from
area S\,111 of the Landolt catalog~\cite{land1:Afanasiev_n}. The
observations were reduced with the DAOPHOT package in the IDL. The
magnitudes are presented in Table~\ref{tab2:Afanasiev_n}.

\subsection{Polarimetry}

Polarimetric observations in the  $V$ filter were carried out on
the SAO~RAS BTA telescope using the SCORPIO-2 focal
reducer~\cite{afan1:Afanasiev_n} and an EEV\,\mbox{42-90} CCD
($4600\!\times\!2048$). A Wollaston prism and a quarter-wave plate
were used as a polarization analyzer. During the $1.^h5$-long
observations on November 6, 2010, eighty images with 60-s
exposures were obtained. A~\mbox{$2\!\times\!2$}~binning and the
restriction of the readout area of the chip to
$711\!\times\!711$~pixels allowed us to reduce the readout time to
5~s. Given the time of rotation of the quarter-wave plate, the
time between the exposures amounted to 70~s.  The size of the
images during that night $d=2''$. An example of the frame is
shown in Fig.~\ref{fig2:Afanasiev_n}, where   the field of the
object can be seen in the ordinary (top) and extraordinary
(bottom) rays. When the quarter-wave plate is rotated, the rays
are interchanged. At the time of the observations, the unit was
not equipped with a suitable mask, hence a slit width of $10''$
was used. Due to limited observing time, we failed to observe the
entire orbital period. The information about the conducted
observations is given in Table~\ref{tab1:Afanasiev_n}.
Figure~\ref{fig5:Afanasiev_n} demonstrates the  variations in
brightness   and circular polarization of the comparison stars and
the reference star. Table~\ref{tab2:Afanasiev_n} presents the mean
values and standard deviations of circular polarization.

Data reduction was conducted in the IDL environment; the
technique,   examples, and formulae can be found
in~\cite{afan2:Afanasiev_n}.

\section{OBSERVATION ANALYSIS}

The obtained  photometric time series have been processed with
V.~P.~Goransky's EFFECT code. Using the Lafler--Kinman method, the
period of the system was refined:
\begin{equation*}
\begin{array}{r}
\label{shiequ1:Afanasiev_n}
\mbox{HJD} =  2455503.2584(\pm 0.001) \\
       + 0.08481(\pm 0.000002)\lefteqn{\,\mbox{E},}
\end{array}
\end{equation*}
where the zero phase corresponds to the minimum brightness time.

Figure~\ref{fig3:Afanasiev_n}a presents the light curve of the
object in the $R_c$ band, folded with our period. The measurement
errors varied from night to night but did not exceed $0.^h3$. They
are not shown in the figure. To illustrate the accuracy of
observations, Fig.~\ref{fig3:Afanasiev_n}b shows the comparison
star brightness variations  obtained on November 2, 2011. The
brightness of the system reaches the maximum at phase
$\varphi=0.65$. The total brightness amplitude of the system
amounts to  $2.^m5$. Let us note two features of the orbital
brightness variations: a deep eclipse and large brightness
fluctuations of up to \mbox{$\Delta m\sim1^{\rm m}$}. The latter
indicates large non-stationarity  of the accretion process.
Because of this, the type of eclipse (full or partial) cannot be
determined.

The $V$-band light curve was a byproduct of our polarimetric
observations (Fig.~\ref{fig4:Afanasiev_n}b). The brightness of the
object was calculated by summation of the intensities of the
ordinary and extraordinary rays in each frame. The brightness
estimates were obtained at intervals of 70~s. Quasiperiodic
variations of
 the system brightness are clearly visible. In our case they occur approximately every
eight minutes with an amplitude of up to $\Delta m\sim0.^m25$.

A comparison of our results with the results
of~\cite{krya1:Afanasiev_n} showed that the scatter of values on
our light curve is twice as large. The light curve
in~\cite{krya1:Afanasiev_n} has an asymmetric shape, which is not
observed for our light curve. This may be due to variations in the
state of the system or inaccurate determination of the period. An
error of  $0.^d0004$  at such a short period may yield a daily
phase shift  by $\varphi=0.083$.

The results of polarimetric observations were folded with our
period. The  circular polarization curve in the   $V$ filter is
shown in Fig.~\ref{fig4:Afanasiev_n}a. Circular polarization of
USNO\,0825 reaches  $-31\%$ at phase \mbox{$\varphi=0.9$},   the
minimum value  is   $-2\%$ at phase \mbox{$\varphi=0.38$}.
Unfortunately, we could not observe throughout the entire period.
The degree of circular polarization can possibly be even higher or
will decrease and will change its sign.

\section{CONCLUSION}

We have presented the results of photometric and polarimetric
observations of \mbox{USNO-A2.0}\linebreak 0825-18396733. Our
polarimetric observations have revealed  that the visible
radiation of the system is highly polarized.  This determines the
system as a polar. Circular polarization in the  $V$ filter
reaches at least   \mbox{$P_V=-31\%$}. The study has refined the
orbital period of the system to \mbox{$P=0.^d08481(2)$} and showed
that the brightness variability amplitude in the  $R_c$ filter is
about~$2.^m5$.   Quasiperiodic   ($P\sim8$~min) brightness
variations of up to $\Delta m\sim0.^m25$ occur in the system. We
have compared  the object with the already studied
V834\,Cen~\cite{crop2:Afanasiev_n} and
\mbox{RX\,J1313.32--3259}~\cite{heyd1:Afanasiev_n} systems, which
are similar in the shape of the brightness  and circular
polarization curves. Both objects at the minimum brightness phases
show weakening of circular polarization. The authors surmise that
there occurs an eclipse of the cyclotron radiation region by an
accretion structure.  At the same time, RX\,J1313.32--3259 shows
the change of the sign of the circular polarization in the blue
range and in the white light   (see Figs.~9~and~10
in~\cite{heyd1:Afanasiev_n}), which indicates the appearance of
the second accretion  region on the visible hemisphere of the
white dwarf.
 As the position angle of linear polarization shows, the main
accretion region is located all the time on the visible hemisphere
of the white dwarf. For a complete analysis of USNO\,0825,
additional polarimetric, spectral, and photometric observations
are required.

\begin{acknowledgments}
The work was supported by the Russian Foundation for Basic
Research (project  No.~14-02-31247). The observations on the 6-m
BTA telescope were carried out with the financial support of the
Ministry of Education and Science of the Russian Federation
(contract no.~14.619.21.0004, project~ID RFMEFI61914X0004).
\end{acknowledgments}

%
% '¯¨á®ª «¨â¥à âãàë
%
\def\saoname{Special Astrophysical Observatory,  Russian Academy of Sciences,
              Nizhnii Arkhyz, 369167 Russia}
\def\saonamer{Специальная астрофизическая обсерватория РАН, Нижний Архыз, 369167
Россия}

%******* SPECIAL SIGNS AND CHARACTERS FOR MATH MODE *******
%
\def\squareforqed{\hbox{\rlap{$\sqcap$}$\sqcup$}}

\def\sq{\ifmmode\squareforqed\else{\unskip\nobreak\hfil
\penalty50\hskip1em\null\nobreak\hfil\squareforqed
\parfillskip=0pt\finalhyphendemerits=0\endgraf}\fi}

\def\sun{\hbox{$\odot$}}

\def\la{\mathrel{\mathchoice {\vcenter{\offinterlineskip\halign{\hfil
$\displaystyle##$\hfil\cr<\cr\sim\cr}}}
{\vcenter{\offinterlineskip\halign{\hfil$\textstyle##$\hfil\cr
<\cr\sim\cr}}}
{\vcenter{\offinterlineskip\halign{\hfil$\scriptstyle##$\hfil\cr
<\cr\sim\cr}}}
{\vcenter{\offinterlineskip\halign{\hfil$\scriptscriptstyle##$\hfil\cr
<\cr\sim\cr}}}}}

\def\ga{\mathrel{\mathchoice {\vcenter{\offinterlineskip\halign{\hfil
$\displaystyle##$\hfil\cr>\cr\sim\cr}}}
{\vcenter{\offinterlineskip\halign{\hfil$\textstyle##$\hfil\cr
>\cr\sim\cr}}}
{\vcenter{\offinterlineskip\halign{\hfil$\scriptstyle##$\hfil\cr
>\cr\sim\cr}}}
{\vcenter{\offinterlineskip\halign{\hfil$\scriptscriptstyle##$\hfil\cr
>\cr\sim\cr}}}}}

\def\degr{\hbox{$^\circ$}}

\def\arcmin{\hbox{$^\prime$}}

\def\arcsec{\hbox{$^{\prime\prime}$}}

\def\utw{\smash{\rlap{\lower5pt\hbox{$\sim$}}}}

\def\udtw{\smash{\rlap{\lower6pt\hbox{$\approx$}}}}

\def\fa{\hbox{$\,.\!\!^{\rm a}$}}

\def\fd{\hbox{$\,.\!\!^{\rm d}$}}

\def\fh{\hbox{$\,.\!\!^{\rm h}$}}

\def\fm{\hbox{$\,.\!\!^{\rm m}$}}

\def\fs{\hbox{$\,.\!\!^{\rm s}$}}

\def\fdg{\hbox{$\,.\!\!^\circ$}}

\def\farcm{\hbox{$\,.\mkern-4mu^\prime$}}

\def\farcs{\hbox{$\,.\!\!^{\prime\prime}$}}

\def\fp{\hbox{$\,.\!\!^{\scriptscriptstyle\rm p}$}}

\def\cor{\mathrel{\mathchoice {\hbox{$\widehat=$}}{\hbox{$\widehat=$}}
{\hbox{$\scriptstyle\hat=$}}
{\hbox{$\scriptscriptstyle\hat=$}}}}

\def\sol{\mathrel{\mathchoice {\vcenter{\offinterlineskip\halign{\hfil
$\displaystyle##$\hfil\cr\sim\cr<\cr}}}
{\vcenter{\offinterlineskip\halign{\hfil$\textstyle##$\hfil\cr\sim\cr
<\cr}}}
{\vcenter{\offinterlineskip\halign{\hfil$\scriptstyle##$\hfil\cr\sim\cr
<\cr}}}
{\vcenter{\offinterlineskip\halign{\hfil$\scriptscriptstyle##$\hfil\cr
\sim\cr<\cr}}}}}

\def\sog{\mathrel{\mathchoice {\vcenter{\offinterlineskip\halign{\hfil
$\displaystyle##$\hfil\cr\sim\cr>\cr}}}
{\vcenter{\offinterlineskip\halign{\hfil$\textstyle##$\hfil\cr\sim\cr
>\cr}}}
{\vcenter{\offinterlineskip\halign{\hfil$\scriptstyle##$\hfil\cr
\sim\cr>\cr}}}
{\vcenter{\offinterlineskip\halign{\hfil$\scriptscriptstyle##$\hfil\cr
\sim\cr>\cr}}}}}

\def\lse{\mathrel{\mathchoice {\vcenter{\offinterlineskip\halign{\hfil
$\displaystyle##$\hfil\cr<\cr\simeq\cr}}}
{\vcenter{\offinterlineskip\halign{\hfil$\textstyle##$\hfil\cr
<\cr\simeq\cr}}}
{\vcenter{\offinterlineskip\halign{\hfil$\scriptstyle##$\hfil\cr
<\cr\simeq\cr}}}
{\vcenter{\offinterlineskip\halign{\hfil$\scriptscriptstyle##$\hfil\cr
<\cr\simeq\cr}}}}}

\def\gse{\mathrel{\mathchoice {\vcenter{\offinterlineskip\halign{\hfil
$\displaystyle##$\hfil\cr>\cr\simeq\cr}}}
{\vcenter{\offinterlineskip\halign{\hfil$\textstyle##$\hfil\cr
>\cr\simeq\cr}}}
{\vcenter{\offinterlineskip\halign{\hfil$\scriptstyle##$\hfil\cr
>\cr\simeq\cr}}}
{\vcenter{\offinterlineskip\halign{\hfil$\scriptscriptstyle##$\hfil\cr
>\cr\simeq\cr}}}}}

\def\grole{\mathrel{\mathchoice {\vcenter{\offinterlineskip\halign{\hfil
$\displaystyle##$\hfil\cr>\cr\noalign{\vskip-1.5pt}<\cr}}}
{\vcenter{\offinterlineskip\halign{\hfil$\textstyle##$\hfil\cr
>\cr\noalign{\vskip-1.5pt}<\cr}}}
{\vcenter{\offinterlineskip\halign{\hfil$\scriptstyle##$\hfil\cr
>\cr\noalign{\vskip-1pt}<\cr}}}
{\vcenter{\offinterlineskip\halign{\hfil$\scriptscriptstyle##$\hfil\cr
>\cr\noalign{\vskip-0.5pt}<\cr}}}}}

\def\leogr{\mathrel{\mathchoice {\vcenter{\offinterlineskip\halign{\hfil
$\displaystyle##$\hfil\cr<\cr\noalign{\vskip-1.5pt}>\cr}}}
{\vcenter{\offinterlineskip\halign{\hfil$\textstyle##$\hfil\cr
<\cr\noalign{\vskip-1.5pt}>\cr}}}
{\vcenter{\offinterlineskip\halign{\hfil$\scriptstyle##$\hfil\cr
<\cr\noalign{\vskip-1pt}>\cr}}}
{\vcenter{\offinterlineskip\halign{\hfil$\scriptscriptstyle##$\hfil\cr
<\cr\noalign{\vskip-0.5pt}>\cr}}}}}

\def\loa{\mathrel{\mathchoice {\vcenter{\offinterlineskip\halign{\hfil
$\displaystyle##$\hfil\cr<\cr\approx\cr}}}
{\vcenter{\offinterlineskip\halign{\hfil$\textstyle##$\hfil\cr
<\cr\approx\cr}}}
{\vcenter{\offinterlineskip\halign{\hfil$\scriptstyle##$\hfil\cr
<\cr\approx\cr}}}
{\vcenter{\offinterlineskip\halign{\hfil$\scriptscriptstyle##$\hfil\cr
<\cr\approx\cr}}}}}

\def\goa{\mathrel{\mathchoice {\vcenter{\offinterlineskip\halign{\hfil
$\displaystyle##$\hfil\cr>\cr\approx\cr}}}
{\vcenter{\offinterlineskip\halign{\hfil$\textstyle##$\hfil\cr
>\cr\approx\cr}}}
{\vcenter{\offinterlineskip\halign{\hfil$\scriptstyle##$\hfil\cr
>\cr\approx\cr}}}
{\vcenter{\offinterlineskip\halign{\hfil$\scriptscriptstyle##$\hfil\cr
>\cr\approx\cr}}}}}

\def\diameter{{\ifmmode\mathchoice
{\ooalign{\hfil\hbox{$\displaystyle/$}\hfil\crcr
{\hbox{$\displaystyle\mathchar"20D$}}}}
{\ooalign{\hfil\hbox{$\textstyle/$}\hfil\crcr
{\hbox{$\textstyle\mathchar"20D$}}}}
{\ooalign{\hfil\hbox{$\scriptstyle/$}\hfil\crcr
{\hbox{$\scriptstyle\mathchar"20D$}}}}
{\ooalign{\hfil\hbox{$\scriptscriptstyle/$}\hfil\crcr
{\hbox{$\scriptscriptstyle\mathchar"20D$}}}}
\else{\ooalign{\hfil/\hfil\crcr\mathhexbox20D}}%
\fi}}

\def\getsto{\mathrel{\mathchoice {\vcenter{\offinterlineskip
\halign{\hfil
$\displaystyle##$\hfil\cr\gets\cr\to\cr}}}
{\vcenter{\offinterlineskip\halign{\hfil$\textstyle##$\hfil\cr\gets
\cr\to\cr}}}
{\vcenter{\offinterlineskip\halign{\hfil$\scriptstyle##$\hfil\cr\gets
\cr\to\cr}}}
{\vcenter{\offinterlineskip\halign{\hfil$\scriptscriptstyle##$\hfil\cr
\gets\cr\to\cr}}}}}

\def\lid{\mathrel{\mathchoice {\vcenter{\offinterlineskip\halign{\hfil
$\displaystyle##$\hfil\cr<\cr\noalign{\vskip1.2pt}=\cr}}}
{\vcenter{\offinterlineskip\halign{\hfil$\textstyle##$\hfil\cr<\cr
\noalign{\vskip1.2pt}=\cr}}}
{\vcenter{\offinterlineskip\halign{\hfil$\scriptstyle##$\hfil\cr<\cr
\noalign{\vskip1pt}=\cr}}}
{\vcenter{\offinterlineskip\halign{\hfil$\scriptscriptstyle##$\hfil\cr
<\cr
\noalign{\vskip0.9pt}=\cr}}}}}

\def\gid{\mathrel{\mathchoice {\vcenter{\offinterlineskip\halign{\hfil
$\displaystyle##$\hfil\cr>\cr\noalign{\vskip1.2pt}=\cr}}}
{\vcenter{\offinterlineskip\halign{\hfil$\textstyle##$\hfil\cr>\cr
\noalign{\vskip1.2pt}=\cr}}}
{\vcenter{\offinterlineskip\halign{\hfil$\scriptstyle##$\hfil\cr>\cr
\noalign{\vskip1pt}=\cr}}}
{\vcenter{\offinterlineskip\halign{\hfil$\scriptscriptstyle##$\hfil\cr
>\cr
\noalign{\vskip0.9pt}=\cr}}}}}

%********** ABBREVIATIONS OF THE OFT-REFERENCED JOURNALS ***********

% *** Астрофизические исследования % русск. версия журнала САО до 1993 г.
\newcommand{\air}{Astrophys. Studies}
% *** Astrophysical Bulletin % англ. версия журнала САО с 2007 г.
\newcommand{\ab}{Astrophysical Bulletin }
% *** Астрофизический бюллетень % русск. версия журнала САО с 2007 г.
\newcommand{\abr}{Астрофизический бюллетень }
% *** Astronomy and Astrophysics
\newcommand{\aaa}{Astron. and Astrophys. }
\newcommand{\aap}{Astron. and Astrophys. }
% *** Astronomy and Astrophys. Supplement Series
\newcommand{\aas}{Astron. and Astrophys. Suppl. }
\newcommand{\aaps}{Astron. and Astrophys. Suppl. }
% *** Astronomy and Astrophysics Review
\newcommand{\aar}{Astron. Astrophys. Rev. }
% *** Astronomical Journal
\newcommand{\aj}{Astron.~J. }
% *** Astrophysical Journal
%\newcommand{\apj}{Astrophys.~J. }
% *** Astrophysical Journal Supplement Series
\newcommand{\apjs}{Astrophys.~J. Suppl. }
% *** Astrophysics and Space Science
\newcommand{\apss}{Astrophys. and Space Sci. }
% *** Annual Review of Astronomy and Astrophys.
\newcommand{\araa}{Annual Rev. Astron. Astrophys. }
% *** Astronomicekij Zhurnal
\newcommand{\azh}{Astron.~Zh. }
% *** Bulletin of the American Astron. Society
\newcommand{\baas}{Bull. Amer. Astron. Soc. }
% *** Bulletin of the Special Astrophysical Observatory % англ. версия до 2007 г.
\newcommand{\bsao}{Bull. Spec. Astrophys. Obs. }
% *** Бюллетень Спец. астрофизич. обсерватории % русск. версия до 2007 г.
\newcommand{\bsaor}{Бюлл. Спец. астрофиз. обсерв. }
% *** Inform. Bul. Var. Stars
\newcommand{\ibvs}{Inform. Bull. Var. Stars }
% *** Journal of Astronomy and Astrophysics
\newcommand{\jaa}{J.~Astron. Astrophys. }
% *** Monthly Notices of the Roy. Astron. Society
\newcommand{\mnras}{Monthly Notices Royal Astron. Soc. }
% *** Publ. of the Astron. Society of Australia
\newcommand{\pasa}{Publ. Astron. Soc. Australia }
% *** Publ. Astronom. Soc. Japan
\newcommand{\pasj}{Publ. Astron. Soc. Japan }
% *** Publ. of the Astron. Society of the Pacific
\newcommand{\pasp}{Publ. Astron. Soc. Pacific }
% *** Astronomy Reports (АЖ)
\newcommand{\arep}{Astronomy Reports }
% *** Astronomy Letters (ПАЖ)
\newcommand{\alet}{Astronomy Letters }
% *** Astronomische Nachrichten
\newcommand{\an}{Astronomische Nachrichten }
% *** Pis'ma v Astronomicekij Zhurnal
\newcommand{\pazh}{Pis'ma Astron. Zh. }
% *** Письма в АЖ
\newcommand{\pazhr}{Письма в АЖ }
% *** Астрон. ж.
\newcommand{\azhr}{Астрон.~ж. }
% *** Soviet Astronomy
\newcommand{\sovast}{Sov. Astron. }
% *** Scientific American
\newcommand{\sca}{Scientific American }
% *** Sky and Telescope
\newcommand{\skytel}{Sky Telesc. }
% *** Space Science Reviews
\newcommand{\spsrev}{Space Sci.~Rev. }
% Revista Mexicana de Astronomia y Astrofisica
%\newcommand{\rmxaa}{Revista Mexicana de Astronom\'{\i}a y Astrof\'{\i}sica}
\newcommand{\rmxaa}{Revista Mexicana Astronom. Astrof\'{\i}s. }
%\newcommand{\nat}{Nature }
% *** Physical Review D
%\newcommand{\prd}{Phys. Rev.~D }
\newcommand{\memsai}{Memorie della Societ\`a Astronomica Italiana}

\newpage

%¨áã­®ª ¢ áâ âìî ¬®¦­® ¢ª«îç¨âì ¯à¨ ¯®¬®é¨ ®ªà㦥­¨ï figure:
\begin{figure*}
\begin{minipage}{0.49\linewidth}
  \vspace{-14pt}
\setcaptionmargin{5mm} \onelinecaptionsfalse
\includegraphics[width=0.925\columnwidth,bb=120 20 490 390,clip]{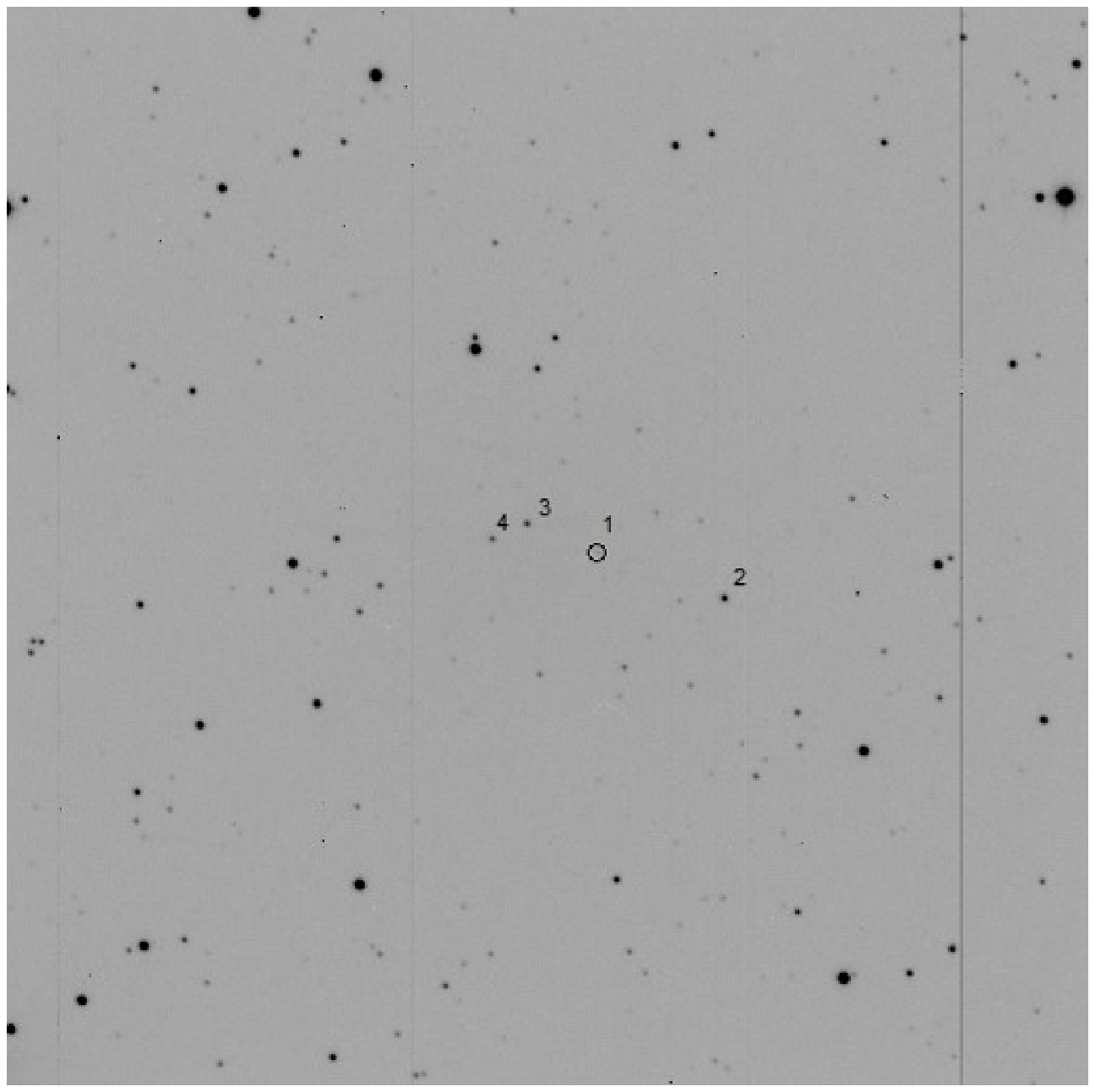}
\caption{An image of the USNO\,0825 region obtained on the
Zeiss-1000 telescope. The numbers denote: ({\it 1})~the object;
({\it 2})~the reference star; ({\it 3,~4})~comparison
stars.}\label{fig1:Afanasiev_n}
\end{minipage}
\begin{minipage}{0.49\linewidth}
 \vspace{1mm}
\setcaptionmargin{5mm} \onelinecaptionsfalse
\includegraphics[width=0.95\columnwidth]{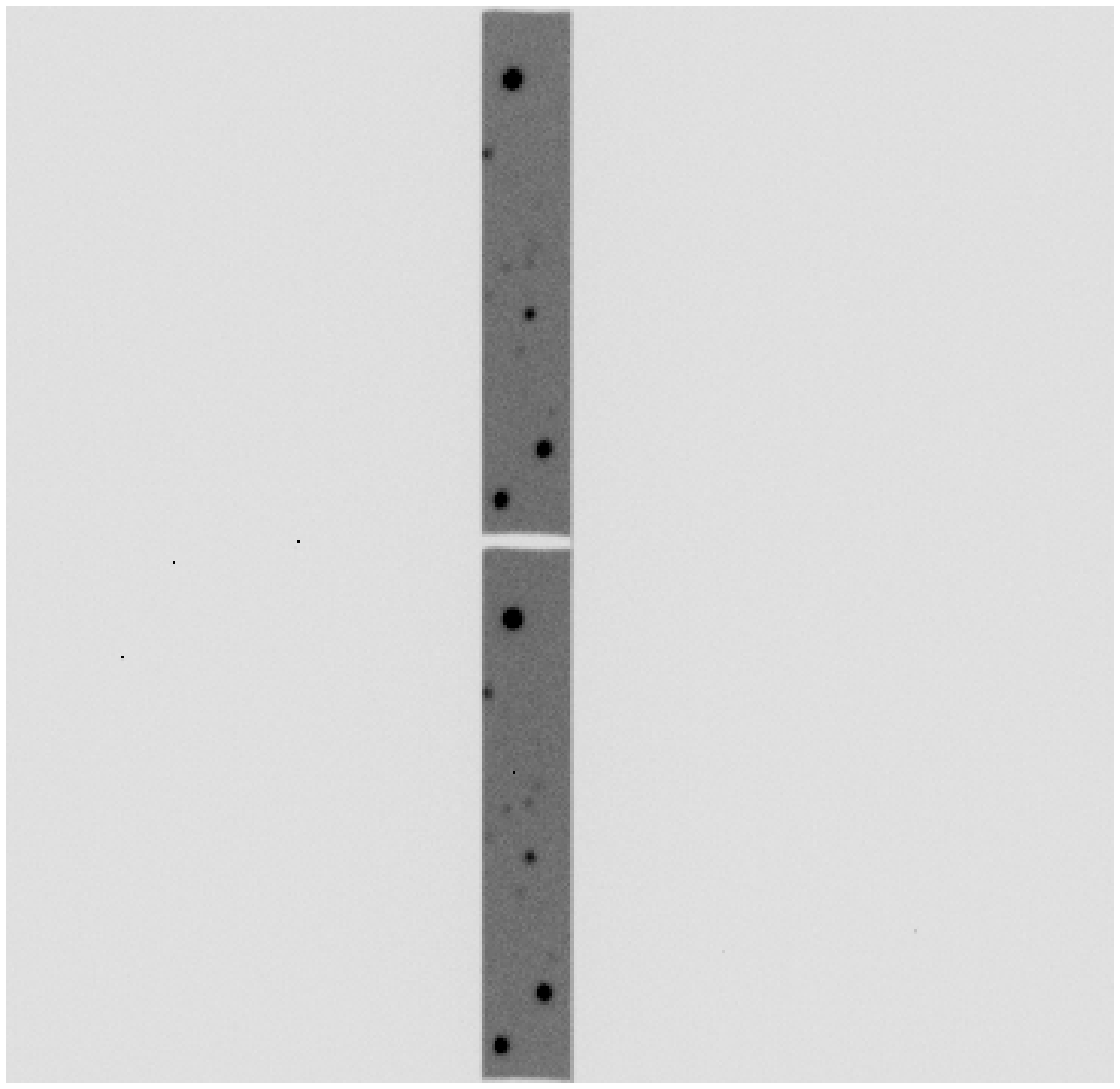}
\caption{An image of the USNO\,0825 region obtained in the
polarimetric mode with SCORPIO-2. Light is divided into the
ordinary (top) and extraordinary (bottom) rays. The numbers
denote: ({\it 1})~the object; ({\it 2})~the reference star; ({\it
3,~4})~comparison stars.}\label{fig2:Afanasiev_n}
\end{minipage}
\begin{minipage}{\linewidth}
 \vspace{5pt}
\setcaptionmargin{0mm} \onelinecaptionstrue
\captionstyle{nonumber}
 \label{tab1:Afanasiev_n} %
\caption{\centerline{{\fontsize{10}{15pt}\selectfont \bf Table~1.}
Observing log of USNO-A2.0\,0825-18396733 observations}}
\addtocounter{table}{1}\addtocounter{figure}{-1}
\medskip
\begin{tabular}{c|c|c|c|c|c|c}
\hline
$ $ & Date   & \multicolumn{2}{c|} {Duration of obs.} & Exposure,  & Number  &  Seeing,  \\
\cline{3-4}
$ $ &  & hours & frac. of the period & s & of  images &  arcsec \\
\hline
             & Oct 6, 2010 & 2.17 & 1.08 & 200 & 29 & 2 \\
Photometry   & Nov 1, 2010 & 2.17 & 1.08 & 100, 120, 150 & 32 & 2.3 \\
             & Nov 2, 2010 & 1.75 & 0.88 & 100 & 34 & 2 \\
\hline
Polarimetry  & Nov 6, 2010 & 1.5 & 0.74 & 60 & 80 & 1.2\\
\hline
\end{tabular}
\end{minipage}
\end{figure*}

\begin{table}[tbp]
\setcaptionmargin{0mm} \onelinecaptionsfalse \captionstyle{normal}
\caption{Magnitudes and  rates of circular polarization of the
reference stars}\label{tab2:Afanasiev_n}
\medskip
\begin{tabular}{c|c|c|r}
\hline
Star & $R_c$, mag & $V$, mag  & \multicolumn{1}{c}{$P_v$, \%}  \\
\hline
{\it 2} & 15.35 & 16.702 & $-0.08\pm0.29$ \\
{\it 3} & $16.05\pm0.04$ &$17.620\pm0.011$ & $-0.17\pm0.77$ \\
{\it 4} & $16.37\pm0.05$ &$17.785\pm0.017$ & $-0.005\pm0.78$ \\
\hline
\end{tabular}
\end{table}

\begin{figure}[tbp!!!]
\setcaptionmargin{5mm} \onelinecaptionsfalse
 \vspace{-1mm}
\includegraphics[width=\columnwidth]{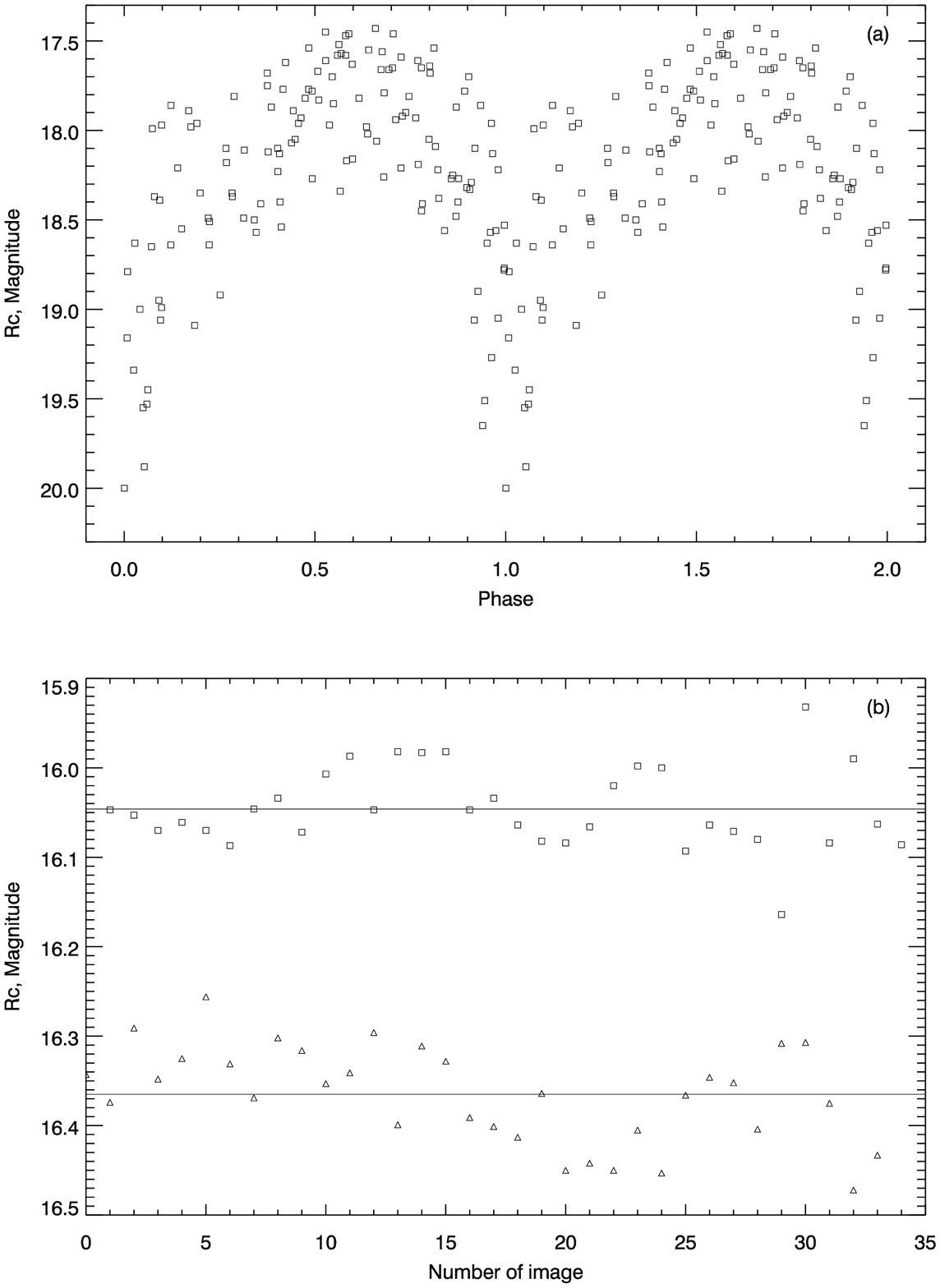}
% \special{psfile=Stars.ps angle=0 hoffset=0 voffset=-30 vscale=83 hscale=85}
\caption{Variations of circular polarization of the comparison
stars and the reference star~(a),  and the   brightness variations
of the comparison stars relative  to the mean value~(b). Triangles
denote the reference star, squares mark comparison star~{\it 3},
diamonds---comparison star~{\it 4} (see
Fig.~1).}\label{fig5:Afanasiev_n}
\end{figure}

\begin{figure}[tbp!!!]
\setcaptionmargin{5mm} \onelinecaptionstrue
 \vspace{-1mm}
\includegraphics[width=\columnwidth]{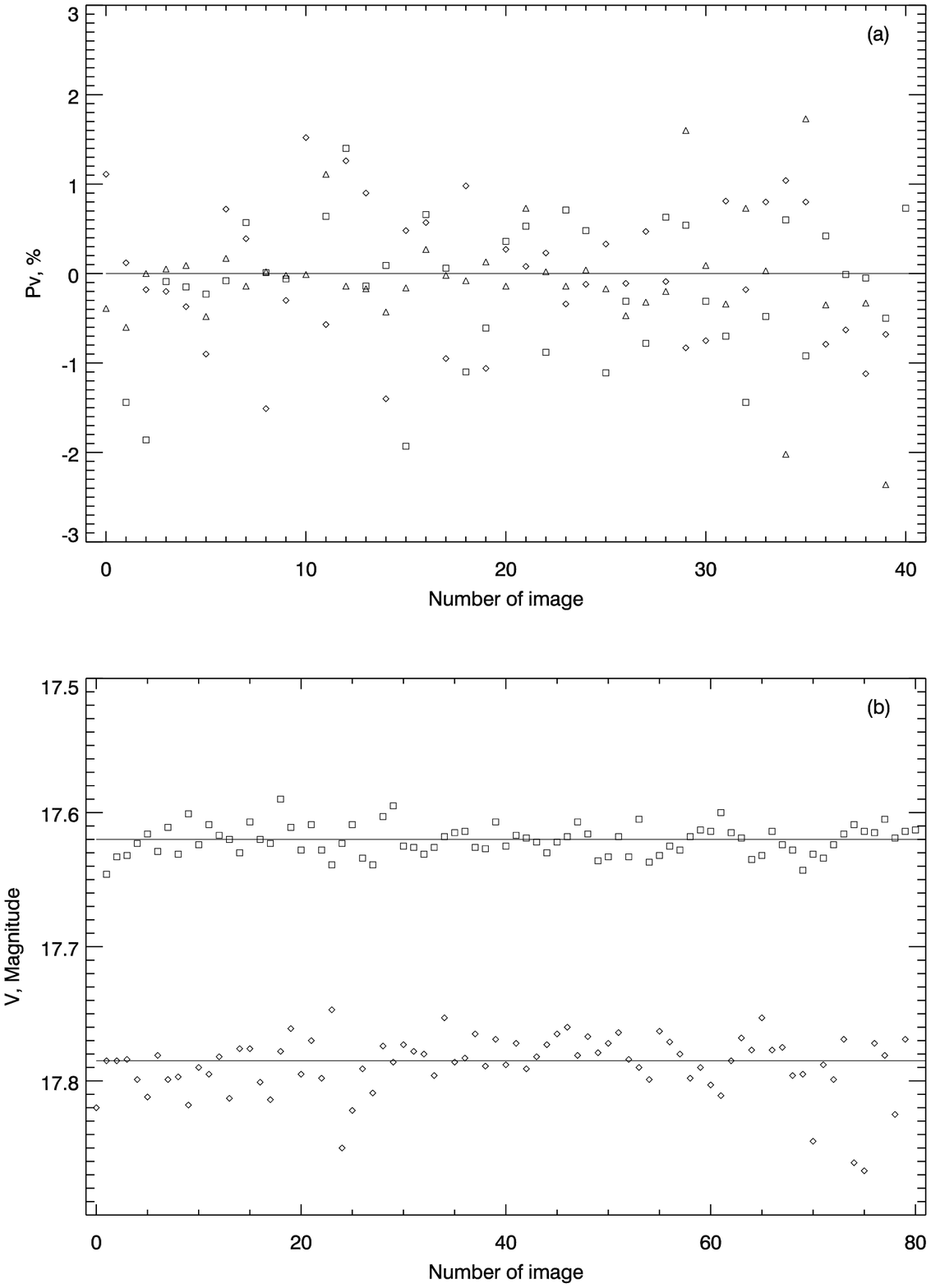}
% \special{psfile=USNO0825.PhotPolV.ps angle=0 hoffset=0 voffset=-30 vscale=83 hscale=85}
\caption{Circular polarization curve~(a) and light curve~(b) of
USNO\,0825 in the $V$ filter.}\label{fig4:Afanasiev_n}
\end{figure}

\begin{figure}[tbp!!!]
\setcaptionmargin{5mm} \onelinecaptionsfalse
 \vspace{-1mm}
\includegraphics[width=\columnwidth]{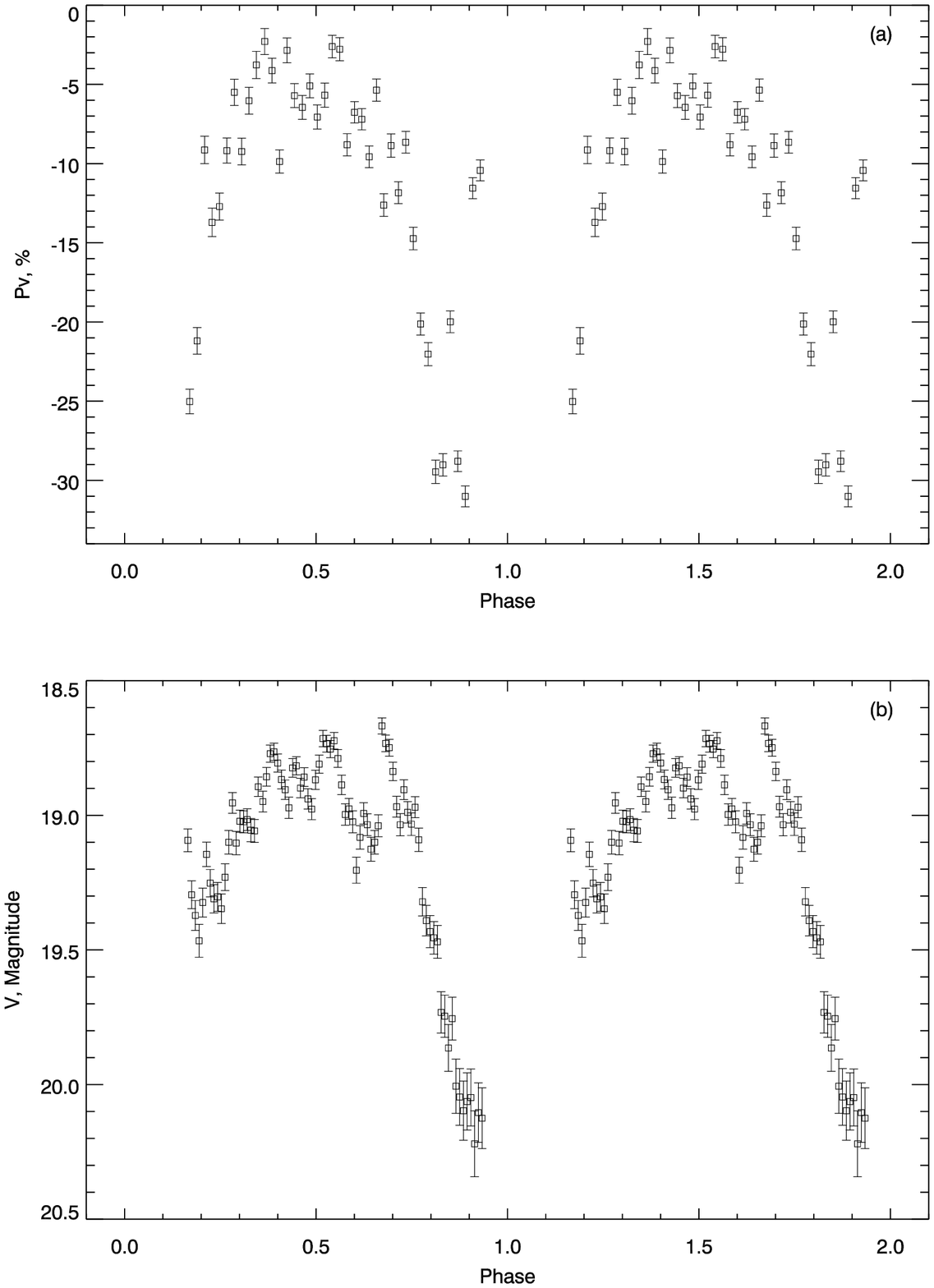}
% \special{psfile=USNO0825.PhotR.all.ps angle=0 hoffset=0 voffset=-30 vscale=83 hscale=85}
\caption{Light curve of USNO\,0825 (a) and brightness variation of
the comparison stars relative to the mean value (b) in the $R_c$
filter.}\label{fig3:Afanasiev_n}
\end{figure}

\end{document}